\documentstyle[psfig]{mn}

\newif\ifAMStwofonts

\ifoldfss
  \ifCUPmtlplainloaded \else
    \NewTextAlphabet{textbfit} {cmbxti10} {}
    \NewTextAlphabet{textbfss} {cmssbx10} {}
    \NewMathAlphabet{mathbfit} {cmbxti10} {} 
    \NewMathAlphabet{mathbfss} {cmssbx10} {} 
  \fi
  \ifAMStwofonts
    \ifCUPmtlplainloaded \else
      \NewSymbolFont{upmath} {eurm10}
      \NewSymbolFont{AMSa} {msam10}
      \NewMathSymbol{\upi}     {0}{upmath}{19}
      \NewMathSymbol{\umu}     {0}{upmath}{16}
      \NewMathSymbol{\upartial}{0}{upmath}{40}
      \NewMathSymbol{\leqslant}{3}{AMSa}{36}
      \NewMathSymbol{\geqslant}{3}{AMSa}{3E}

    \fi
  \fi
\fi 

\ifnfssone
  \newmathalphabet{\mathit}
  \addtoversion{normal}{\mathit}{cmr}{m}{it}
  \addtoversion{bold}{\mathit}{cmr}{bx}{it}
  \newmathalphabet{\mathbfit} 
  \addtoversion{normal}{\mathbfit}{cmr}{bx}{it}
  \addtoversion{bold}{\mathbfit}{cmr}{bx}{it}
  \newmathalphabet{\mathbfss} 
  \addtoversion{normal}{\mathbfss}{cmss}{bx}{n}
  \addtoversion{bold}{\mathbfss}{cmss}{bx}{n}
  \ifAMStwofonts
    \ifCUPmtlplainloaded \else
      \UseAMStwoboldmath
      \makeatletter
      \new@mathgroup\upmath@group
      \define@mathgroup\mv@normal\upmath@group{eur}{m}{n}
      \define@mathgroup\mv@bold\upmath@group{eur}{b}{n}
      \edef\UPM{\hexnumber\upmath@group}
      \new@mathgroup\amsa@group
      \define@mathgroup\mv@normal\amsa@group{msa}{m}{n}
      \define@mathgroup\mv@bold\amsa@group{msa}{m}{n}
      \edef\AMSa{\hexnumber\amsa@group}
      \makeatother
      \mathchardef\upi="0\UPM19
      \mathchardef\umu="0\UPM16
      \mathchardef\upartial="0\UPM40
      \mathchardef\leqslant="3\AMSa36
      \mathchardef\geqslant="3\AMSa3E
    \fi
  \fi
\fi 

\ifnfsstwo
  \DeclareMathAlphabet{\mathbfit}{OT1}{cmr}{bx}{it}
  \SetMathAlphabet\mathbfit{bold}{OT1}{cmr}{bx}{it}
  \DeclareMathAlphabet{\mathbfss}{OT1}{cmss}{bx}{n}
  \SetMathAlphabet\mathbfss{bold}{OT1}{cmss}{bx}{n}
  \ifAMStwofonts
    \ifCUPmtlplainloaded \else
      \DeclareSymbolFont{UPM}{U}{eur}{m}{n}
      \SetSymbolFont{UPM}{bold}{U}{eur}{b}{n}
      \DeclareSymbolFont{AMSa}{U}{msa}{m}{n}
      \DeclareMathSymbol{\upi}{0}{UPM}{"19}
      \DeclareMathSymbol{\umu}{0}{UPM}{"16}
      \DeclareMathSymbol{\upartial}{0}{UPM}{"40}
      \DeclareMathSymbol{\leqslant}{3}{AMSa}{"36}
      \DeclareMathSymbol{\geqslant}{3}{AMSa}{"3E}
    \fi
  \fi
\fi 

\ifCUPmtlplainloaded \else
  \ifAMStwofonts \else 
    \def\upi{\pi}
    \def\umu{\mu}
    \def\upartial{\partial}
  \fi
\fi

\title{A flattening in the Optical Light Curve of SN 2002ap}
\author[S. B. Pandey et al.]
      {S. B. Pandey$^{1}$\thanks{E-mail: shashi@upso.ernet.in}, G. C. Anupama$^2$, R. Sagar$^{1,2}$
, D. Bhattacharya$^{3}$, 
\newauthor D. K. Sahu $^{4}$ and J. C. Pandey$^{1}$ \\
        $^{1}$State Observatory, Manora Peak Nainital 263129, India\\
        $^{2}$Indian Institute of Astrophysics, Bangalore 560034, India\\
        $^{3}$Raman Research Institute, Bangalore 560080, India\\     
        $^{4}$Center for Research \& Education in Science \& Technology, Hosakote, 562114, India}
\date{Accepted --------- .
      Received --------- ;
                         }

\pagerange{\pageref{firstpage}--\pageref{lastpage}}
\pubyear{1994}

\begin{document}

\maketitle

\label{firstpage}

\begin{abstract}
We present the $UBVR_cI_c$ broad band optical photometry of the Type Ic 
supernova SN 2002ap obtained during 2002 February 06 -- March 23
in the early decline phases and also later on 2002 15 August. Combining these 
data with the published ones, the general light curve development is studied. 
The time and luminosity of the peak brightness and the 
peak width are estimated. There is a flattening in the optical light
curve about 30 days after the $B$ maximum.
The flux decline rates before flattening are 0.127$\pm$0.005,
0.082$\pm$0.001, 0.074$\pm$0.001, 0.062$\pm$0.001 and 0.040$\pm$0.001
mag day$^{-1}$ in $U$, $B$, $V$, $R_c$ and $I_c$ passbands respectively, while 
the corresponding values after flattening are about 0.02 mag day$^{-1}$ in all 
the passbands. The maximum brightness of SN 2002ap $M_V = - 17.2$ mag, is comparable 
to that of the type Ic 1997ef, but fainter than that of the type Ic hypernova SN 1998bw. 
 The peak luminosity indicates an ejection of $\sim$ 0.06 M$_{\odot}$ ${}^{56}$Ni mass.

We also present low-resolution optical spectra obtained during the early phases. 
 The SiII absorption minimum indicates that the photospheric velocity decreased from $\sim$
 21,360 km s$^{-1}$ to $\sim$ 10,740 km s$^{-1}$ during a period of $\sim$ 6 days.

\end{abstract}


\section{Introduction}

The supernova SN 2002ap ($\alpha_{2000}$ = $01^{h}36^{m}23^{s}.92 $;
$\delta_{2000}$ = $+15^{\circ}$ $45^{\prime}$ $13^{\prime\prime}.3$) was
discovered on 2002 January 29.4 UT by Y. Hirose at $V \sim 14.5$ mag, in the
outer region of the nearby spiral M74 (Nakano et al.\ 2002). Low resolution 
spectra of SN 2002ap obtained during 2002 January 30--31 (Kinugasa et al. 
2002a,b; Meikle et al. 2002; Filippenko \& Chornock 2002) 
were similar to those of type Ic supernovae. However, the spectral 
features were found to be extremely broad resembling the type Ic `hypernovae'
SN 1997ef and SN 1998bw (Filippenko 1997; Nomoto et al. 2002; Mazzali et al. 2002). 

At a distance of 7.3~Mpc, SN 2002ap being the nearest hypernova discovered to 
date, was a good target for a detailed monitoring, and has been subject to multi-wavelength 
observations. In addition to optical observations, it has been observed in the 
X-ray (Sutaria et al.\ 2002), radio (Berger, Kulkarni \& Chevalier 2002, Sutaria et al.\ 2002) 
and in the infrared (Mattila \& Meikle 2002). Evidence for a high velocity asymmetric 
explosion has been indicated by spectropolarimetric observations (Kawabata et 
al.\ 2002; Leonard et al.\ 2002; Wang et al.\ 2002). 

The broad spectral features in the spectrum and the evolution
of Si II spectral line indicate a high expansion velocity ($\sim$ 30000
km s$^{-1}$), which supports hypernova model for the SN 2002ap (Kinugasa et al.
2002b; Meikle et al. 2002; Gal-Yam et al. 2002a,b; Filippenko \& Chornock
2002). The spectropolarimetric observations during the early phase indicate a
similarity with SN 1998bw (Leonard et al.\ 2002). Modeling of the optical
observations indicate SN 2002ap as an energetic event, with an explosion
energy of $\simeq 4-10\times 10^{51}$~ergs (Mazzali et al. 2002). However, the 
radio observations seem to indicate that SN 2002ap an ordinary type Ic 
supernova, without a jet (Berger et al. 2002). A study of the 
$UBVRIH_{\alpha}K$ images of galaxy M74 obtained several years prior to the 
discovery of SN 2002ap (Smartt et al. 2002) resulted in a non-detection of the
progenitor.

The spectral similarity to SN 1998bw, the possible link between very energetic
supernova and gamma ray bursts and the lack of substantive data on rare type Ic SN events
make SN 2002ap very important object to study in
detail. We have therefore carried out dense temporal multi-colour optical
photometric observations during the early phases. These in combination with the published
data are used to study the development of the optical light curve. 
Our observations are the first to indicate a flattening in the light curve of the
SN 2002ap about 30 days after the B maximum. 
We also present low resolution optical spectra obtained during the early phases.
The details of both photometric and spectroscopic observations are presented in the 
next section, while the development of the light curves, spectral and other properties of the 
SN 2002ap are discussed in the remaining part of the paper.

\section{Observations and Data Reduction}
\subsection{Photometry}

During the early phase, optical $UBVR_cI_c$ observations of SN2002ap were 
carried out for 32 days between 2002 February 06 and March 23 from the State 
Observatory, Nainital, India using a 2048$\times$2048 pixel$^2$ CCD system 
attached at the f/13 Cassegrain focus of the 104 cm Sampurnanand reflector. 
One pixel of the CCD chip corresponds to a square of $\sim$ 0.38 arcsec while the 
entire chip covers
a field of 13$ ^ \prime \times 13^ \prime$ on the sky. The gain and read out noise of
the CCD camera are 10 $e^-/ADU$ and 5.3 $e^-$ respectively. The supernova
was observed in the $BVR_cI_c$ bands during most of the nights, while $U$ band
observations could be obtained only for 3 nights. Exposure times for most
of the $U$, $B$, $V$, $R_c$ and $I_c$ band images were 300, 100, 100, 60 and 60
s respectively. The observations of SN 2002ap could not be carried out during 2002 April
 to July due to its proximity to the sun as well as rainy sky conditions in India.
SN2002ap was however observed on 2002 August 15 in $VR_cI_c$ bands using the SITe $1024 \times 1024$ CCD 
system at the Cassegrain focus of the 2~m Himalayan Chandra Telescope at the 
Indian Astronomical Observatory, Hanle, India. Two frames, each with exposure 
time of 420 s, were obtained in $V$; three frames, with exposure times of 300, 420 and 
540 s were obtained in $R_c$ and three frames, each with an exposure time of 300 s
were obtained in $I_c$. Several bias and twilight flat frames were
obtained by the both CCD cameras to calibrate the supernova images using standard techniques. Data
reduction was carried out using IRAF\footnote{IRAF is distributed by National 
Optical Astronomy Observatories, USA.} and MIDAS softwares. For photometric
calibration, comparison stars 1 and 2 of Gal-Yam et al. (2002b)
were observed along with SN~2002ap.  The magnitudes of SN 2002ap and the
comparison stars were estimated using a fixed aperture photometry. 
Since SN 2002ap is located in the outer
region of M74, the contamination due to the light of the galaxy is expected
to be negligible in the measurements (Smartt, Ramirez \& Vreeswijk 2002).
The zero-point accuracy of the $UBVR_cI_c$ magnitudes is of order 0.02
mag.  In order to use the observations made during non-photometric sky
conditions, differential photometry has been adopted, assuming that the errors
introduced due to colour differences between comparison stars and SN
2002ap are much smaller than the zero point errors. 

Following the calibration provided by Henden et al (2002), we have used
$U = 14.10$ mag, $B = 13.84$ mag, $V = 13.06$ mag, $R_c$ = 12.61 mag, 
$I_c$ = 12.14 mag for star 1 and $U = 14.40$ mag, $B = 14.33$ mag, 
$V = 13.69$ mag, $R_c = 13.32$ mag, $I_c = 12.95$ mag for comparison star 2.
The differences in the measured $BVR_cI_c$ magnitudes of the two
comparison stars do not show any secular trend during our observations.
The standard deviation of these differences are 0.013, 0.026, 0.033 and
0.020 mag in the $B$, $V$, $R_c$ and $I_c$ passbands respectively, indicating 
the level of accuracy of our photometry. The photometric errors in the $U$ band
are higher, of order 0.04 mag. As the comparison stars are 
located in much cleaner environments than SN 2002ap, farther from the 
nucleus of M74, their photometry can be considered to be more accurate than 
that of SN 2002ap. The $UBVR_cI_c$ magnitudes of the SN 2002ap estimated based 
on our observations are tabulated in Table 1 along with errors.

\subsection{Spectroscopy}

Spectroscopic observations of SN 2002ap were made from the Vainu Bappu 
Observatory, Kavalur, India, on 2002 February 3.61, 4.60, 5.60 and 10.59 UT. CCD 
spectra, in the range 3800--8000~\AA, were obtained using the OMR spectrograph
on the 2.3m Vainu Bappu Telescope. The spectra were obtained at a resolution
of 600 on February 3 and 4, while the spectra of February 5 and 10 were 
obtained at a resolution of 1200. All spectra were obtained with a narrow slit 
of 2 arcsec width aligned at $0^\circ$~PA. Spectrophotometric standard
HD 19445 was observed on Feb 3, while Hiltner 600 was observed on all other
nights. Both standards were observed with a slit of 5~arcsec width. 

All spectra were bias subtracted, flat-field corrected, extracted and
wavelength calibrated in the standard manner using the IRAF
reduction package. The spectra were corrected for instrumental response,
 telluric absorption lines and brought to a relative flux scale using the 
spectrophotometric standard observed on the same 
night. The fluxes have been brought to an absolute flux scale
using zero points derived from $UBVR_cI_c$ photometry.

\section{Distance and reddening} 

Interstellar NaI D1 absorption features due to the gas in
M74 indicate a reddening of $E(B-V) = 0.008\pm 0.002$ mag due to the
host galaxy (Klose, Guenther \& Woitas 2002). The Galactic
extinction estimated from Schlegal et al. (1998) is $E(B-V) = 0.07$ mag.
Thus the galactic extinction clearly dominates the total extinction of 
$E(B-V) = 0.08$ mag. The extinction in different passbands
corresponding to this reddening are calculated using the relations given
by Cardelli, Clayton \& Mathis (1989) for normal interstellar reddening.
The values thus obtained are listed in Table 2.
 
There is no Cepheid distance determination to M74. Most of the SN 2002ap
studies have used the mean distance of 8 Mpc for M74 group derived by
Sharina et al. (1996), except Smartt et al. (2002) who used a distance
of 7.3 Mpc for M74. As the difference between these two values
corresponds to a difference of 0.2 mag in distance modulus and does not
have a significant effect on the conclusions drawn in the paper, we have
used a distance of 7.3 Mpc for M74. In the light of the above, we conclude
that distance modulus of M74 has an uncertainty of about 0.2 mag while
the total reddening is uncertain by $\pm 0.01$ mag.

\section{Description of Spectra}

\begin{figure}
\psfig{file=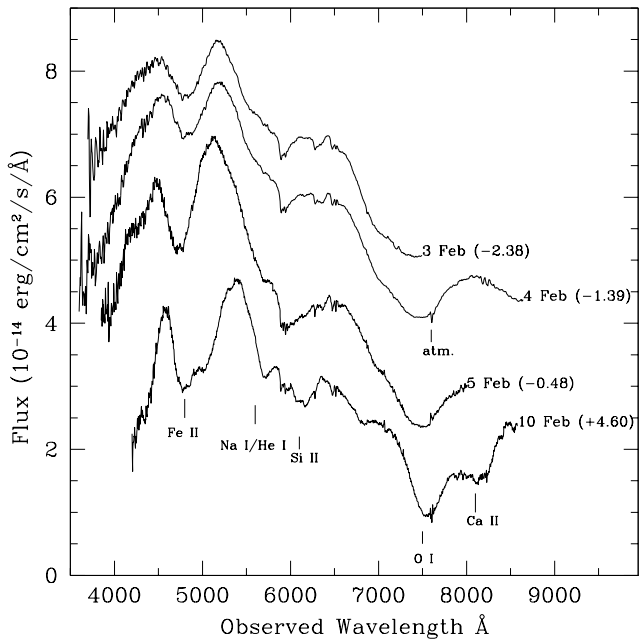,height=9.0cm,width=9.0cm}
\caption{Optical spectra of SN2002ap. The dates of observations along with 
the days relative to B maximum are indicated. In order to avoid  overlap, 
an offsets of 4.5, 3.5 and 1.7 in flux units, has been added to the 
spectra of Feb 3.61, 4.60 and 5.60 respectively.}
\end{figure}

Figure 1 shows the spectra of SN 2002ap. Offsets of 4.5, 3.5, and 1.7 in flux
units have been added to the spectra of Feb 3.61, 4.60 and 5.60 respectively.
The lines in the spectra have been identified following Mazzali et al. 
(2002). The spectra presented here fill in the gap in the pre-maximum phase
observations presented by Mazzali et al. (2002) and Kinugasa et al. (2002b). 
Following Kinugasa et al. (2002b), we estimate the photospheric velocity using
the Si II 6347, 6371 \AA\AA (6355~\AA) absorption minimum. Not much variation
is detected in the photospheric velocity for the spectra of Feb 3.61, 4.60
and 5.60, corresponding to $-2.38$, $-1.39$, $-0.48$ days from $B$ maximum. The
average of all three spectra gives a value of $21,360\pm 2000$~km s$^{-1}$, 
corresponding to day $-1.39$ from $B$ maximum, or 6.7 days since explosion.
Date of explosion is assumed to be 2002 Jan 28.9 UT 2002 (Mazzali et al. 2002). The 
velocity decreased to a value of $10,740\pm 1500$~km s$^{-1}$ on Feb 10.59,
4.60 days after $B$ maximum, or 12.69 days since explosion. These estimates are
in excellent agreement with the velocity evolution estimate by Kinugasa et al. 
(2002b) for SN 2002ap.

\begin{figure}
\psfig{file=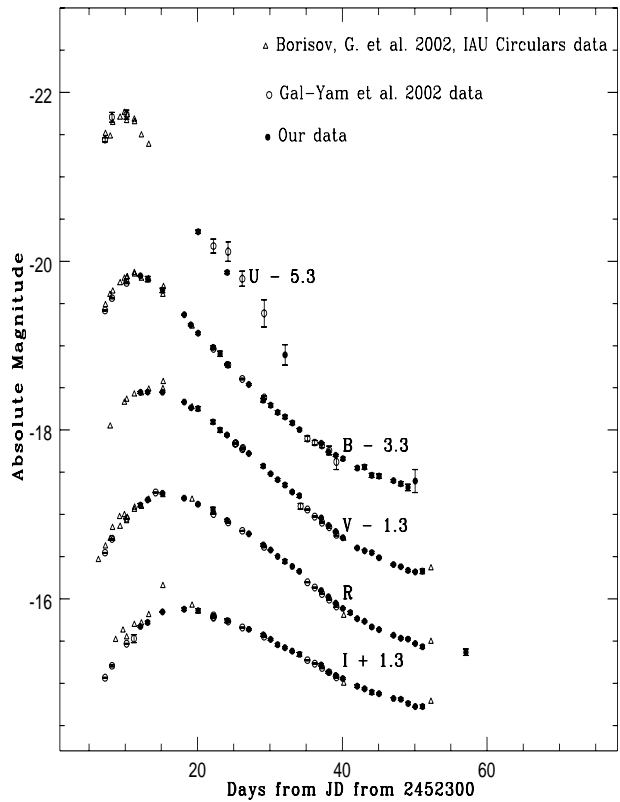,height=12.0cm,width=9.0cm}
\caption{UBV$R_cI_c$ light curve of SN 2002ap. The light curves are offset by a
constant value on the magnitude scale as indicated in the plot.} 
\end{figure}
    
\begin{table*}

{\bf Table 1.}~$UBVR_c$ and $I_c$ magnitudes of SN 2002ap along with errors, Julian date 
and mid UT of observations are listed.
\scriptsize
\begin{center}
\begin{tabular}{cccccccc} \hline\hline
Date (UT)     & Time in JD & $I_c$ (mag)  &  $R_c$ (mag)  & $V$ (mag) & $B$ (mag) & $U$ (mag) & \\ \hline
Feb 06.62 & 2452312.12  & 12.46$\pm$0.02&12.39$\pm$0.04&12.42$\pm$0.03&13.12$\pm$0.02&$-$&                 \\  
Feb 07.61 & 2452313.11  & 12.42$\pm$0.03&12.33$\pm$0.03&12.42$\pm$0.03&13.16$\pm$0.03&$-$& 				\\
Feb 09.66 & 2452315.16  & 12.29$\pm$0.02&12.25$\pm$0.03&12.42$\pm$0.03&13.29$\pm$0.03&$-$&				\\
Feb 12.66 & 2452318.15  & 12.26$\pm$0.03&12.31$\pm$0.03&12.53$\pm$0.03&13.58$\pm$0.02&$-$&				\\ 
Feb 13.58 & 2452319.08  & $-$      &$-$       &12.60$\pm$0.03&13.70$\pm$0.02&$-$&                               \\ 
Feb 14.57 & 2452320.07  & 12.28$\pm$0.04&12.38$\pm$0.03&12.61$\pm$0.04&13.80$\pm$0.02&14.64$\pm$0.02&           \\  
Feb 16.64 & 2452322.14  & $-$      &12.45$\pm$0.04&12.77$\pm$0.03&13.97$\pm$0.02&$-$&                              \\
Feb 17.63 & 2452323.13  & $-$      &$-$       &12.86$\pm$0.03&14.04$\pm$0.03&$-$&                               \\ 
Feb 18.58 & 2452324.08  & 12.40$\pm$0.03&12.58$\pm$0.03&12.92$\pm$0.03&14.17$\pm$0.02&15.12$\pm$0.02&     			\\ 
Feb 21.58 & 2452327.08  & 12.50$\pm$0.02&12.74$\pm$0.03&13.14$\pm$0.03&14.41$\pm$0.02&$-$&     				\\ 
Feb 23.57 & 2452329.07  & 12.56$\pm$0.02&12.87$\pm$0.03&13.29$\pm$0.03&14.59$\pm$0.02&$-$&    		\\   
Feb 24.60 & 2452330.10  & 12.62$\pm$0.02&12.93$\pm$0.03&13.38$\pm$0.03&14.65$\pm$0.02&$-$&     				\\  
Feb 25.56 & 2452331.06  & 12.68$\pm$0.03&13.00$\pm$0.03&13.45$\pm$0.03&14.74$\pm$0.02&$-$&     				\\ 
Feb 26.58 & 2452332.08  & 12.72$\pm$0.02&13.06$\pm$0.04&13.52$\pm$0.03&14.79$\pm$0.02&16.10$\pm$0.12&    \\
Feb 27.58 & 2452333.08  & 12.76$\pm$0.02&13.12$\pm$0.03&13.60$\pm$0.03&14.86$\pm$0.02&$-$&\\
Feb 28.59 & 2352334.10  & 12.79$\pm$0.03&13.18$\pm$0.03&13.64$\pm$0.03&14.94$\pm$0.02&$-$ &    				\\
Mar 03.58 & 2452337.08  & 12.92$\pm$0.02&13.41$\pm$0.03&13.91$\pm$0.03&15.10$\pm$0.02&$-$& \\
Mar 04.58 & 2452338.08  & 13.01$\pm$0.02&13.48$\pm$0.03&13.99$\pm$0.03&15.20$\pm$0.02&$-$ &   		 		 \\ 
Mar 05.58 & 2452339.08  & 13.04$\pm$0.02&13.56$\pm$0.03&14.07$\pm$0.03&15.25$\pm$0.02&$-$  &   				\\ 
Mar 06.57 & 2452340.07  & 13.08$\pm$0.02&13.62$\pm$0.03&14.14$\pm$0.03&15.28$\pm$0.02& $-$		&	\\   
Mar 07.60 & 2452341.01  & 13.17$\pm$0.03&13.67$\pm$0.04&$-$       &$-$ 	     &$-$			&\\ 
Mar 08.57 & 2452342.07  & 13.20$\pm$0.02&13.74$\pm$0.03&14.26$\pm$0.03&15.40$\pm$0.02&  $-$    			&\\  
Mar 09.57 & 2452343.07  & 13.24$\pm$0.03&13.77$\pm$0.03&14.30$\pm$0.03&15.38$\pm$0.03&  $-$    	&	\\   
Mar 10.57 & 2452344.07  & 13.26$\pm$0.02&13.84$\pm$0.03&14.32$\pm$0.03&15.48$\pm$0.03&  $-$    	&		\\  
Mar 11.57 & 2452345.07  & 13.32$\pm$0.02&13.87$\pm$0.03&14.38$\pm$0.03&15.49$\pm$0.03&  $-$    	&		\\  
Mar 13.57 & 2452347.07  & 13.33$\pm$0.02&13.94$\pm$0.03&14.46$\pm$0.03&15.55$\pm$0.02&  $-$    	&		\\ 
Mar 14.58 & 2452348.08  & 13.38$\pm$0.02&$-$        &14.48$\pm$0.03&15.58$\pm$0.02&  $-$    	&		\\ 
Mar 15.58 & 2452349.08  & $-$       &13.97$\pm$0.03&14.53$\pm$0.03&15.62$\pm$0.04&  $-$    	&		\\ 
Mar 16.58 & 2452350.08  & 13.41$\pm$0.02&13.98$\pm$0.03&14.55$\pm$0.03&15.55$\pm$0.14&  $-$   	&			\\
Mar 17.58 & 2452351.08  & 13.41$\pm$0.03&14.03$\pm$0.03&14.54$\pm$0.04&$-$        &  $-$   	&			\\ 
Mar 18.58 & 2452352.08  & $-$       &14.07$\pm$0.04&$-$       &$-$  	     &$-$		&	\\
Mar 23.57 & 2452357.07  & $-$       &14.14$\pm$0.05&$-$       &$-$           &$-$    	&			\\ 
Aug 15.90 & 2452502.40  & 16.37$\pm$0.03&16.46$\pm$0.03&17.32$\pm$0.03 &$-$   & $-$ &  \\
\hline 
\hline
\end{tabular}
\end{center}
\end{table*}

\section{The $UBVR_cI_c$ light curve and its development}

In order to produce temporally dense light curves in $UBVR_cI_c$, we have
combined our observations with published ones, taking data from
Gal Yam et al. (2002b), Borisov et al.  (2002), Yoshii et
al. (2002), Matohara et al., (2002), Riffeser et al. (2002), Hornock \& Lelekovice (2002), 
 Henden et al.(2002a). The frequency distribution of 
the data taken from the literature is N($U, B, V, R_c, I_c$) = (18, 28, 24, 33, 27). 
The combined $UBVR_cI_c$ light curves during the early phase are shown in Fig. 2. A good agreement
can be clearly seen in the different data sets indicating that, within
observational errors, they are on the same photometric scale. 

\subsection{First 60 days}

Fig. 2 indicates that the light curve of SN 2002ap rose fast but declined slowly in an exponential
manner. The light curve appears to have flattened
about 30 days after $B$ maximum in all the passbands. Using a smooth cubic
spline interpolation between the data points, the peak brightness and its time
of occurrence are were estimated for all the passbands the values thus obtained are tabulated
in Table 2. These values are consistent with, but
more accurate than those presented by Gal-Yam et al. (2002b).  

The general nature of the light curve of SN 2002ap is similar to that of SN
1998bw (Galama et al. 1998). The light curve of SN 2002ap and its characteristics
such as the time of maxima, peak width and the decline slopes are compared with those 
of the hypernova SN 1998bw and the normal type Ic supernova SN 1994I 
(Richmond et al. 1996). The values of these parameters are listed in Table 2. 
In all cases, the flux peaks first at
shorter wavelengths and then moves progressively towards longer wavelengths
as expected in energetic explosions such as supernovae. The time differences
between the maxima at $B$ and $I_c$ filters are about 5.5, 3.5 and 3.2 days
for SN 2002ap, SN 1998bw and SN 1994I respectively. 
We define the peak width $\Delta d_{0.25}$ as the width of the light curve at a
level 0.25 magnitude below the maximum. The value of $\Delta d_{0.25}$ therefore can be 
considered
as a representative of rise time in the light curve. From Table 2, it can be seen that the peak
width ($\Delta d_{0.25}$) of SN 2002ap is smaller than that of SN 1998bw but larger than that
of SN 1994I. The $\Delta d_{0.25}$ value is larger at longer wavelengths
for all the three SNe and decreases gradually towards the shorter
wavelengths. As the absolute magnitude at the time of peak brightness of
SN 2002ap is fainter than that of SN 1998bw and SN 1994I, the amount of
radio active matter ejected in the SN 2002ap explosion is probably smaller 
than those in SN 1998bw and SN 1994I.

\begin{figure}
\psfig{file=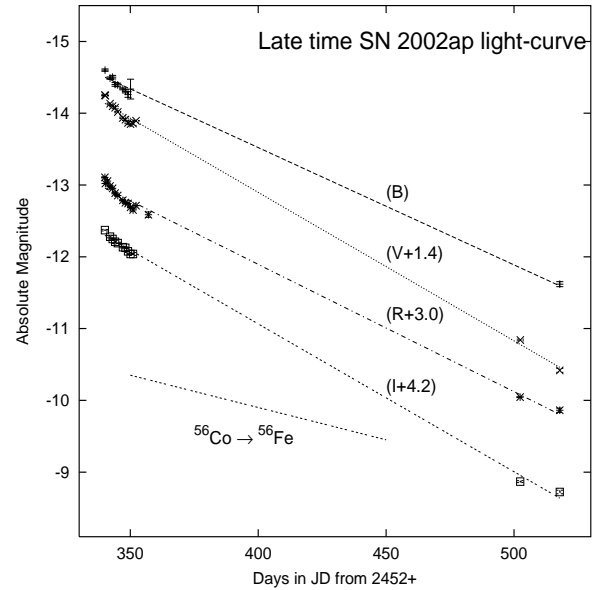,height=8.0cm,width=11.0cm,angle=270}
\caption{ Late time light curve of SN 2002ap. The light curves are offset by a
constant value in the magnitude scale, as indicated in the plot. For comparison
, slope of the ${}^{56}$Co $\rightarrow$ ${}^{56}$Fe decay light curve is also shown.} 
\end{figure}

The value of $\Delta d_{0.25}$ peaks in different bands depends on the total 
ejected mass and the explosion energy (Iwamoto et al. 1998, 
Iwamoto et al. 2000).
The smaller value of $\Delta d_{0.25}$ as seen in the case of SN 1994I
means that this supernova ejected a relatively lower mass, producing a 
relatively thin envelope, which allowed the inner core photons
to escape more quickly. Ejection of more massive envelopes increases the
diffusion time and leads to a broader peak at light maximum.
This appears to be true in the case of SN 1997ef and SN 1998bw. We
therefore expect the mass ejected in SN 2002ap to be more compared to 
SN 1994I but slightly less compared to SN 1998bw. This is in agreement 
with the estimate of the ejected mass determined by Mazzali et al. (2002). 
Following Richmond et al. (1996), we also estimate $\Delta m_{15}$, the 
magnitude of SN 2002ap 15 days after the outburst maximum, and 
find the values to be intermediate between SN 1998bw and SN1994I (see Table 2). 
It can be seen from Table 2 that the values of $\Delta m_{15}$ indicate an
anti-correlation between the decline rate and $\Delta$d$_{0.25}$, in the sense 
that a larger decline rate corresponds to a shorter rise time.

Figure 2 indicates a flattening in the slope of the light curve of
SN 2002ap after $\sim$ JD 2542340 in all the passbands.
A similar flattening has been observed in the light curves of SN 1998bw and 
SN 1994I also. However the time of occurrence of this flattening, as measured from
the $B$ maximum, is different for the three SNe under discussion. It is about 
30 days for SN 2002ap, $\sim$ 35 days for SN 1994I but $\sim$ 40 days for SN 1998bw.
The flux decline rates ($\alpha_1$) prior to the flattening for SN 2002ap are wavelength dependent, with 
values $\sim$ 0.12, 0.08, 0.07, 0.06 and 0.04 mag day$^{-1}$ in $U$, $B$, $V$, $R_c$ and 
$I_c$ passbands respectively. A similar trend is also found for SN 1998bw and SN 1994I (see table 2).
 Thus the early flux decline values are faster at shorter 
wavelengths, indicative of expansion of the ejecta OE. cooling photosphere. 

\subsection{Light curve after flattening}

Figure 3 shows the late time $BVR_cI_c$ light curve from JD 2542340 to 
JD 2542520, following the occurrence of the flattening.
The slopes of the flattened part ($\alpha_2$) of the light curves are 
$\sim$ 0.02 mag day$^{-1}$ in $B$, $V$, $R_c$ and $I_c$ bands corresponding to a value 
of $\sim$ 35 days for the half life decay time of the radio active nuclei. 
A similar decline rate was observed in the case of SN 1998bw also (Patat et al.\ 2001).
A comparison of $\alpha_1$ and $\alpha_2$ values indicates that the general
nature of all the three SNe are similar for the phases under consideration.  
The $\Delta$m$_{15}$ and $\alpha_1$ values indicate that the early flux decline rate
 depends upon wavelength, being fastest at shorter wavelengths. On the other hand 
the values of $\alpha_2$ appear to be independent of wavelength. 

Further observations are planned for a detailed study of the late phase light curve development.

\begin{table*}
\scriptsize
{\bf Table 2.}~ A comparison of the light curve characteristic parameters of 
SN 2002ap, SN 1998bw 
and SN 1994I. Peak JD dates denotes the time of maximum brightness in different filters. 
$\Delta$d$_{0.25}$ is defined as the width of light curve (in days) at a level of 0.25 magnitude 
below maximum.  $\Delta$m$_{15}$ is defined as decline in magnitude in 15 days after maximum. 
$\alpha_1$ and $\alpha_2$ are the flux decline slopes in mag day$^{-1}$ before 
and after the flattening respectively. 
For comparison we have used data points from Galama et al.(1998), 
McKenzie \& Schaefer (1999) and 
Sollerman et al. (2000) for SN 1998bw and from Richmond et al. (1996) for 
SN 1994I.
\begin{center}
\begin{tabular}{ccccccc} \hline
Object & Parameters   &   $U$   &  $B$  &  $V$  &  $R_c$  &  $I_c$  \\ \hline \hline

{\bf SN 2002ap}&Peak JD date&09.77$\pm$0.26&11.49$\pm$0.25&13.91$\pm$0.37&15.90$\pm$0.22&17.24$\pm$0.53\\
&from 2452300+ &&&&&\\
&App. Magnitude&13.26$\pm$0.02&13.12$\pm$0.01&12.38$\pm$0.01&12.28$\pm$0.01&12.22$\pm$0.03 \\
            &&&&&&\\
&Absolute Magnitude&-16.48$\pm$0.02&-16.57$\pm$0.01&-17.22$\pm$0.01&-17.25$\pm$0.01&-17.23$\pm$0.03\\
            &&&&&&\\
&Adopted Extinction&0.427&0.369&0.279&0.209&0.135\\
            &&&&&&\\
&$\Delta$m$_{15}$ &--&0.99&0.87&0.73&0.47 \\
            &&&&&&\\
&$\Delta$d$_{0.25}$ &5.2&7.9&10.6&13&15.3 \\
            &&&&&&\\
&$\alpha_1$&0.127$\pm$0.005&0.082$\pm$0.001&0.074$\pm$0.001&0.062$\pm$0.001&0.040$\pm$0.001 \\
&$\alpha_2$&..... &0.0163$\pm$0.0001&0.0206$\pm$0.0002&0.0177$\pm$0.0004&0.0207$\pm$0.0002 \\

            &&&&&&\\
{\bf SN 1998bw}&Peak JD date& 2.9$\pm$0.2& 3.8$\pm$0.2& 5.7$\pm$0.2& 6.7$\pm$0.2&7.3$\pm$0.3 \\
&from 2450940+&&&&&\\
&Absolute Magnitude&-19.16$\pm$0.10&-18.88$\pm$0.05&-19.35$\pm$0.05&-19.36$\pm$0.05&-19.27$\pm$0.05\\
       &&&&&&\\
&$\Delta$m$_{15}$ &--&0.56&0.41&0.25&0.18 \\
            &&&&&&\\
&$\Delta$d$_{0.25}$&8.9&11.3&13.3&14.4&18.0 \\
            &&&&&&\\
&$\alpha_1$&0.106$\pm$0.007&0.077$\pm$0.003&0.061$\pm$0.001&0.050$\pm$0.001&0.042$\pm$0.002 \\
&$\alpha_2$&.....&0.015$\pm$0.001&0.019$\pm$0.001&0.020$\pm$0.001&0.018$\pm$0.001 \\

            &&&&&&\\
{\bf SN 1994I}&Peak JD date&49.5$\pm$0.1& 50.02$\pm$0.25&51.86$\pm$0.09&52.54$\pm$0.11&53.24$\pm$0.12\\
        &from 2449400+&&&&&\\
&Absolute Magnitude&-17.99$\pm$0.84&-17.68$\pm$0.73&-18.09$\pm$0.58&-17.99$\pm$0.48&-17.78$\pm$0.38\\
       &&&&&&\\
&$\Delta$m$_{15}$ &--&2.07&1.74&1.46&1.08 \\
            &&&&&&\\
&$\Delta$d$_{0.25}$ &--&6.8&7.3&8.3&8.8 \\
            &&&&&&\\
&$\alpha_1$&--&0.125$\pm$0.011&0.122$\pm$0.002&0.107$\pm$0.002&0.084$\pm$0.002 \\
&$\alpha_2$&--&0.021$\pm$0.003&0.024$\pm$0.001&0.021$\pm$0.001&0.023$\pm$0.001 \\
    
\hline
\end{tabular}
\end{center}
\end{table*}

\begin{figure}
\psfig{file=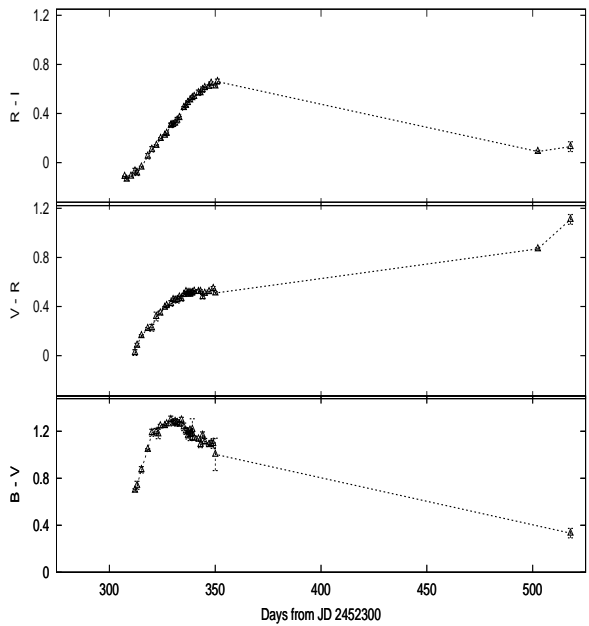,height=8.0cm,width=8.0cm,angle=270}
\caption{Variation of different colours with time for SN 2002ap}
\end{figure}

\section{Colour Curves of SN 2002AP}

The optical colour curves for SN 2002ap are shown in Figure 4. The earliest
measured colours of SN 2002ap are considerably redder than is typical for
most supernovae of type Ic. However, they are similar to those of SN 1994I 
(Richmond et 
al.\ 1996). All the colours, ($B-V$), ($V-R$) and ($R-I$), redden from 
discovery to about JD 2452330. The ($B-V$) colour then turns blue, until the 
end of the observations. 
This turning point most probably occurs when the supernova makes the transition from being
optically thick to optically thin. The ($V-R$) colour on the
other hand, continue to redden reaching about 1.1 mag at the end of our observations.
The $(R-I)$ colour also shows a trend similar to the $(B-V)$ colour. 

The red $(V-R)$, and blue $(R-I)$ colours around 200 days since $B$ maximum 
can be explained as due to a contribution from strong emission lines to the $R$ 
band.

\section{Bolometric Light Curve of SN 2002ap}

Figure 5 shows the UVOIR bolometric light curve of SN 2002ap, obtained
by interpolating and adding contributions from $U$, $B$, $V$, $R_c$, $I_c$, $J$,
 $H$ and $K$ bands. The magnitudes were converted to flux using calibrations by 
Fukugita et al. (1995) for the $U$, $B$, $V$, $R_c$ and $I_c$ bands and by 
Bessell \& Brett (1988) for the $J$, $H$, and $K$ bands. 
The $U$ band contribution has been ignored after JD 2452330, 
as it falls to negligible values.  For the IR bands, we estimate the
contribution to the total flux to be $\sim$ 20\% at early epochs 
(around JD 2452307) and $\sim$ 40\% after JD 2452330, as 
indicated by Mazzali et al. (2002).  
For a comparison, the slopes of $ {}^{56}$Ni $\rightarrow$ $ {}^{56}$Co and 
$ {}^{56}$Co $\rightarrow$ $ {}^{56}$Fe decay curves are also shown in Fig. 5.
The bolometric light curve indicates that after reaching a maximum the flux 
declines exponentially with a decay rate of $0.055\pm 0.001$ mag day$^{-1}$. 

Including the late time observations of JD\ 2452502 and JD\ 2452517 also, the bolometric light
curve indicates a slope of $0.0199\pm 0.0004$ mag day$^{-1}$ beyond JD\,2452340,
assuming a 40\% IR contribution to the total bolometric flux for the late time also, 
as assumed by Mazzali et al. (2002) for the early time.
The late time decline is steeper than that expected from ${}^{56}$Co $\rightarrow$ ${}^{56}$Fe decay rate
, indicating a leakage of $\gamma$-rays from the SN envelope (Sollerman et al.\ 1998; Patat et al.\ 2001) .

\subsection{${}^{56}$Ni mass ejected}

The amount of $^{56}$Ni mass ejected can be estimated using the peak luminosity (Arnett 1982).
Using the peak luminosity of $\sim$ 1.6$\times$10$^{42}$ erg/sec, we estimate the
amount of $^{56}$Ni ejected to be $\sim$ 0.06 M$_{\odot}$. This value is 
consistent with an estimate of 0.07$\pm$0.02 M$_{\odot}$, obtained by Mazzali
et al.\ (2002) using hydrodynamic model fits to the light curves. 

\begin{figure}
\psfig{file=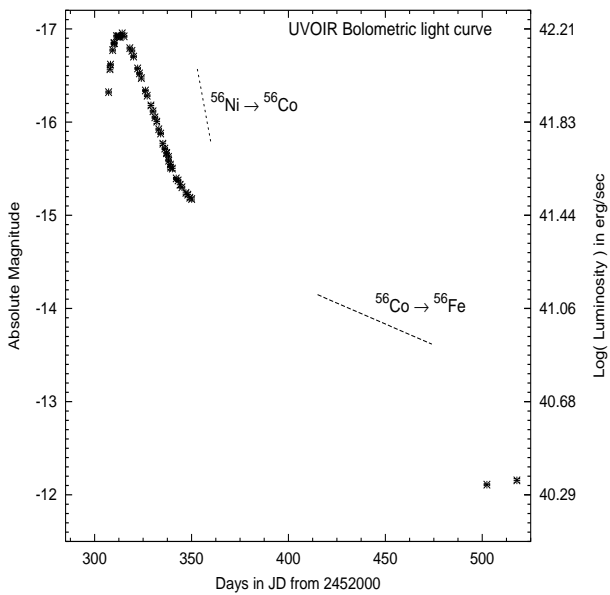,height=8.0cm,width=10.0cm,angle=270}
\caption{UVOIR bolometric light curve of SN 2002ap. For a comparison, the decay slopes of 
${}^{56}$Ni $\rightarrow$ ${}^{56}$Co and ${}^{56}$Co $\rightarrow$ ${}^{56}$Fe are also shown.} 
\end{figure}

\section{Conclusions}

We present dense temporal optical photometric data of SN 2002ap
during 2002 February 06 to March 23. $UBVR_cI_c$ photometric light curves of SN
2002ap have been studied by combining our data with those published by
others. The broad band photometric observations taken up to about 40 days 
after the peak brightness in $B$ indicate that the flux undergoes an
exponential decline similar to that observed in other SN Ic. We are the 
first to report a flattening in the optical light curves at JD 2452340, 
about 30 days after the $B$ maximum.

The supernova luminosity follows an exponential decline with a relatively faster
rate up to JD 2452340. The flux decline rates are
0.127$\pm$0.005, 0.082$\pm$0.001, 0.074$\pm$0.001, 0.062$\pm$0.001 and
0.040$\pm$0.001 mag day$^{-1}$ in $U$, $B$, $V$, $R_c$ and $I_c$ filters respectively.
This indicates a clear dependence of flux decline rate on wavelength, being
faster at shorter wavelengths. On the other hand the flux decline rates
appear to be wavelength independent after flattening with a value of about
0.02 mag day$^{-1}$. The photospheric velocities determined by us are $\sim$ 21,360
 and $\sim$ 10,740 km s$^{-1}$ about $-2.38$ days before and $+4.60$ days after the
$B$ maxima. 

We present the bolometric light curve which illustrates the decay of total 
luminosity of the supernova. The peak luminosity estimated yields the value
of $^{56}$Ni mass ejected to be 0.06 M${_\odot}$. 
The measured early decline rate of the bolometric light curve is slower than 
that expected from ${}^{56}$Ni $\rightarrow$ ${}^{56}$Co decay, but not inconsistent
with that expected from a mixture of iron-peaked nuclei 
(Colgate and McKee 1969). However as seen from the synthetic light curve in
Mazzali et al. (2002), this could also represent an optical depth effect.
 Late time bolometric light curve decline, steeper than that of ${}^{56}$Co $\rightarrow$ ${}^{56}$Fe
 decay is may be due to $\gamma$-rays leakage from the SN envelope.

We intend to monitor SN 2002ap further, in order to study the nature of 
its flux decline at late phases.

\section*{Acknowledgments}
We are thankful to the anonymous referee for valuable comments.      
We also thank the assistants at the Vainu Bappu Telescope for obtaining some of
the spectra. One of us (SBP) is grateful to Dr. Vijay Mohan for his help with
the photometric data reduction and for several useful discussions.

\appendix

\bsp

\label{lastpage}

\end{document}